# Minimum sample size for detection of Gutenberg-Richter's b-value

Yavor Kamer[1, 2]

14.07.2013

[1]Swiss Seismological Service, Institute of Geophysics, ETH Zürich, Switzerland
[2]Chair of Entrepreneurial Risks, Department of Management, Technology and Economics, ETH Zürich, Switzerland

**Abstract:**

In this study we address the question of the minimum sample size needed for distinguishing between Gutenberg-Richter distributions with varying b-values at different resolutions. In order to account for both the complete and incomplete parts of a catalog we use the recently introduced angular frequency magnitude distribution (FMD). Unlike the gradually curved FMD, the angular FMD is fully compatible with Aki's maximum likelihood method for b-value estimation. To obtain generic results we conduct our analysis on synthetic catalogs with Monte Carlo methods. Our results indicate that the minimum sample size used in many studies is strictly below the value required for detecting significant variations.

**Introduction:**

Since its introduction in the 90s [*Wiemer and Benoit*, 1996], the spatial mapping of *Gutenberg and Richter* [1954] law's (GR) b-value has become an increasingly popular method for physical interpretation of various geological structures, faulting regimes [*Wiemer and Wyss*, 1997, 2002] and volcanoes [*Mcnutt et al.*, 2004]. Several studies have related spatial b-value variations to shear stress [*Schorlemmer et al.*, 2005] and pore pressure variations [*Bachmann et al.*, 2012]. *Schorlemmer and Wiemer* [2005] even suggested predictability of location and area of future ruptures on the basis of b-value anomaly. These findings have often been confronted by arguments maintaining that these b-value variations are often observed due to under sampling, magnitude errors and non-homogenous detection capabilities [*Shi and Bolt*, 1982; *Frohlich and Davis*, 1993; *Kagan*, 1999, 2002, 2010; *Amorese et al.*, 2010].

In this study we investigate what is the minimum sample size required to distinguish between varying b-values at different resolutions. Obviously as the sample size increases the variance on the estimations will decrease and this will in turn allow for a better resolution (ability to differentiate between smaller b-value variations). Under such a simplistic view calculating the confidence intervals with respect to the sample size for a single b-value estimation (i.e *b*=1) would be sufficient. Assuming a normal distribution of errors and a statistical significance limit of %5 would indicate that two b-values should be separated by at least 4 standard deviations, which would be equivalent to the resolution. However this assumption is greatly undermined because (1) the error function of b-value estimation is asymmetric (i.e the probability of overestimation is greater than underestimation) (2) the error function is dependent on the b-value itself (i.e higher b-values have broader confidence intervals). Moreover studies investigating the b-value uncertainties usually consider only the events above the detection threshold (*Mc*) and thus do not take into account variance of this parameter [*Shi and Bolt*, 1982]. In a recent study *Amorese et al.* [2010] considered the entire magnitude range using the cumulative normal distribution function to describe the incomplete part as suggested by *Ogata and Katsura* [1993]. However their results are limited to the cases of *b*=1 and *b*=0.7. In this study we investigate the detection threshold for different resolutions on the interval of 0.5≤b≤1.5 which is a commonly reported range in many studies. We prefer to use the angular FMD recently introduced by *Mignan* [2012] which allows for increased computational efficiency and simple analytical derivations. We first give a brief overview of the angular FMD and its advantages over the gradually curved FMD. We then derive analytical implications of the angular FMD shape for detecting different b-values. Lastly we present results of Monte Carlo simulations and the tabulated minimum sample sizes required for detecting anomalous high and low b-values under varying resolutions.

**Gradually Curved vs Angular FMD**

In their study *Ogata and Katsura* [1993] proposed an FMD model to describe the entire observed magnitude range. The model assumes that the observations are governed by two processes acting over the whole magnitude range: (1) the GR law $\lambda_0(M,b)$ and (2) the detection rate which is modeled by a cumulative normal distribution function $q(M,\mu,\sigma)$. Together these two functions result in an intensity function $\lambda = \lambda_0(m)q(m)$ (normalized number of events at *m*) with 3 free parameters given in equation (1) where $\mu$ is the magnitude of %50 detection, $\sigma$ relates to the detection drop-off and *b* is the exponent of the GR law.

$$\lambda(m|\mu,\sigma,b) = 10^{-bm} \int_{-\infty}^{m} \frac{1}{\sqrt{2\pi}\sigma} \exp\left(-\frac{x-\mu}{2\sigma^2}\right) dx \tag{1}$$

The shape of $\lambda_0(M,b)$ and $q(M,\mu,\sigma)$ are given in Figure 1a. The main limitation of this FMD model is that the probability of not detecting an event extends to the region of large magnitudes. This results in the incomplete part affecting the exponent of the complete part, an effect which is rather pronounced in small sample sizes. This is also why the b-value is incompatible with *Aki's* [1965] maximum likelihood estimate.

In order to tackle these issues we shall use the angular FMD introduced by *Mignan* [2012] which models the incomplete part by a detection drop-off described with a power law (shown in Figure 1b). The functional form of the proposed model is given in Equation (2). $\lambda(m)$ is the total number of events smaller (for the incomplete part) or greater (for the complete part) than any magnitude *m* for a magnitude of completeness *Mc*. $10^{ai}$ and $10^{ac}$ are the total number of incomplete and complete events, *b* is the exponent of the GR law obtained from the complete part, *k* is the detection rate drop-off exponent and $k' = k-b$. The effect of varying each parameter is shown in Figure 1c.

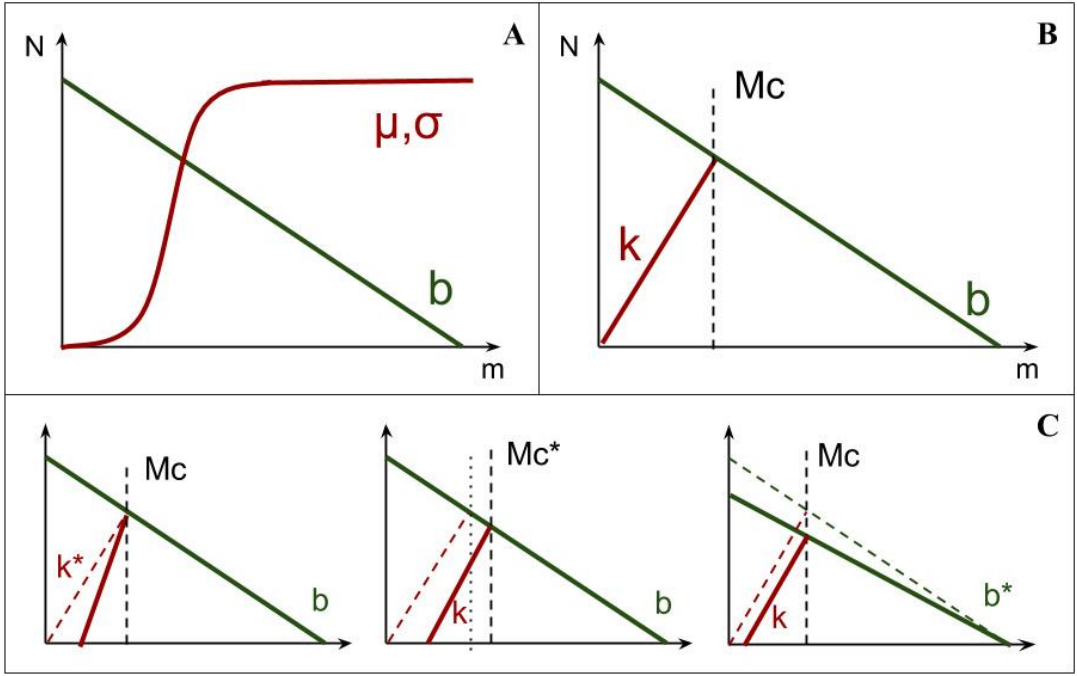

**Figure 1.** Functional shape of (A) a gradually curved FMD, (B) an angular FMD. (C) Effect of varying each of the 3 angular FMD parameters, the varying parameter is marked with an asterisk.

$$\lambda(m \mid Mc, b, k') = \begin{cases} 10^{a_i - k'(Mc - m)}, & m < Mc \\ 10^{a_c - b(m - Mc)}, & m \geq Mc \end{cases} \quad (2)$$

The detection rate is effective only in the incomplete region (m<Mc) and thus does not affect the exponent of the complete part. This allows the use of *Aki's* [1965] maximum likelihood estimate and reduces the computational cost. In the same study *Mignan* [2012] shows that the angular FMD outperforms its predecessor and is a more likely model for local limited sized samples.

**Analytical Implications of the Angular FMD**

In this section we present some analytical derivations using the angular FMD which have implications for the a-value (productivity parameter) of the GR law. We show that these implications affect the detection resolution for a given fixed sample size.

For a sample size with a total number of events $N_t$, the events with a magnitude larger than *Mc* will constitute the complete part ($N_c$) and the rest will be incomplete ($N_i = N_t - N_c$). Since the angular FMD is a piecewise probability density function (PDF) integrating to 1, each part is described by a separate function (i.e there is no overlap or transition), thus we can integrate these parts separately. To obtain the PDFs we normalize $\lambda_0$ with $N_t$ which results in the cumulative density function and then we take derivative with respect to *m*. The integral of each PDF is thus equal to the completeness and incompleteness ratio of the sample.

$$\int_{Mc}^{\infty} \frac{1}{N_t} 10^{a_c - b(m - Mc)} b \log(10) = \frac{N_c}{N_t}$$

$$10^{a_c} = N_c$$

(3)

$$\int_{-\infty}^{Mc} \frac{1}{N_t} 10^{a_i - k'(Mc - m)} k' \log(10) = \frac{N_i}{N_t}$$

$$10^{a_i} = N_i$$

(4)

Since the FMD is continuous, for $m=Mc$ both PDFs should be equal. We substitute $10^{a_i}$ and $10^{a_c}$ in equation (5):

$$\frac{1}{N_t} 10^{a_c} b \log(10) = \frac{1}{N_t} 10^{a_i} k' \log(10)$$

$$\frac{N_c b}{N_t} = \frac{N_i k'}{N_t}$$

$$\frac{b}{k'} = \frac{N_i}{N_c}$$

(5)

For the inverse problem this result implies that if the magnitudes of the events are known the angular FMD has only two free parameters. If $b$ and $k'$ (through $k$) are chosen then $Mc$ is bound to be located at the abscissa of their intersection. Conversely if $b$ and $Mc$ are chosen $k'$ is bound since $\lambda_0(Mc, b, k')$ integrates to one. Next we investigate what these results suggest for the ratio of complete to incomplete events in different b-value regimes. Solving $N_t$ from equation (5) we get $N_t = N_c(k' + b)/k'$, also by definition we have $k' = k - b$, substituting these into equation (3) we obtain the ratio of completeness as:

$$\int_{Mc}^{\infty} \frac{1}{N_t} 10^{a_c - b(m - Mc)} b \log(10) = \frac{N_c}{N_t}$$

$$\frac{N_c k'}{N_c(k' + b)} = 1 - \frac{b}{k}$$

(6)

To see the implications of these result let's assume a stable local station network with a constant detection rate drop-off exponent $k=3$. Equation (6) implies that for a constant seismicity rate and a given time length a region with a b-value of ($b=1$) will have only %66 of its events complete while a region with a b-value of ($b=0.5$) will have %83 completeness. This would mean that any b-value mapping approach which uses just the complete part with fixed sample sizes and finds different b-values is necessarily looking at different time intervals. This is also intuitive since a low b-value region produces larger earthquakes which have a higher probability of being detected while a high b-value region has more small events on the border of detection.

**Implications for b-value detection resolution**

In the previous section we showed that unlike its gradually curved predecessor the angular FMD is compatible with *Aki'*s [1965] b-value estimator and also allows us to account for the change of completeness induced by *b* variation. In this section we present how these new results affect the minimum sample size required for detection of different b-values at varying resolutions.

Previous studies [*Wiemer and Wyss*, 2002; *Felzer*, 2006] tried to partially address this issue by calculating confidence intervals for only *b*=1 with synthetics considering only the complete part. For this study we consider the entire magnitude range by creating synthetic catalogs using the angular FMD with the parameters *k*=3, *Mc*=2 and varying *b* between 0.5≤b≤1.5 (a commonly reported interval) at an increment of Δ*b*=0.02. The choice of *k*=3 is compatible with observed values in local seismicity catalogs such as Parkfield (*k*≈2.9) and also supported by the findings of *Mignan* [2012] for whole California. The *Mc* parameter has no effect on the results. To take into account the magnitude uncertainty observed in real catalogs, the simulated magnitudes are binned into ΔM =0.1. For each b-value increment we simulate 1000 catalogs with the number of complete events starting from 30 and increased gradually to 3000. We obtain the *b* and *Mc* parameters of the angular FMD which maximize the likelihood of each catalog through grid searching over *Mc*. We remind the reader that during this procedure *b* is estimated using *Aki'*s [1965] maximum likelihood formula and *k* is obtained from equation (5), which are corrected for binned magnitudes [*Bender*, 1983]. In Figure 2 we plot the %5 and %95 confidence intervals for the whole b-value range with an increment of Δ*b*=0.2. For small sample size we observe that higher b-values have broader confidence intervals while for lower b-values these are considerably narrower. In order to put this into perspective we can make an analogy of a thermometer (shown at both sides of Figure 2); at small sample sizes such a thermometer would measure lower temperatures more precisely then higher ones. As the sample size is increased the precision will become uniform across the measurement range. Inspecting Figure 2 we can conclude that for a resolution of Δ*b*=0.2 the required complete size is ~700. This is the abscissa of the intersection of the lower bound (5%) of *b*=1.5 and upper bound (95%) of *b*=1.3. Smaller sample sizes may still allow differentiating between b-values on the lower end; however the higher end of the b spectrum will be erroneous due to the crossover of the confidence intervals.

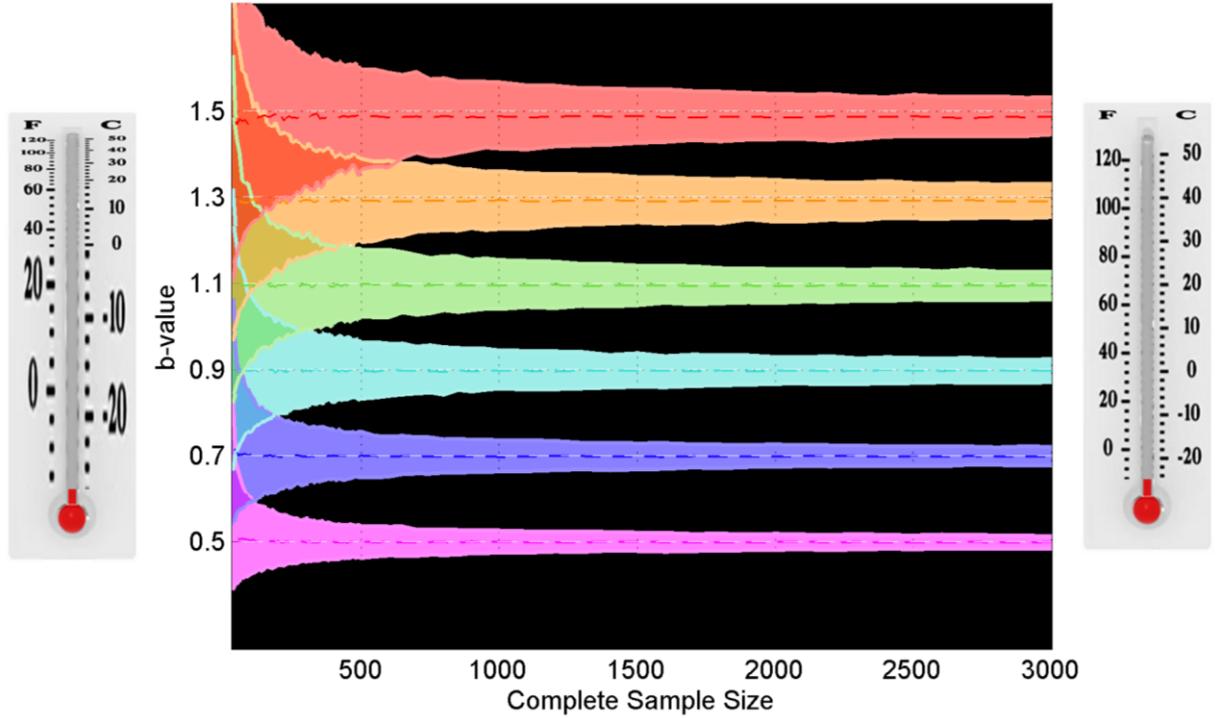

**Figure 2.** Confidence intervals of 5% and 95% for Δb=0.2. The analogous thermometers for small and large sample sizes are given at left and right side of the plot.

In Figure 2 we assume that each b-value region has the same number of complete events, however according to the results obtained in the previous section this is not the case. If the seismicity rate of the region is constant this implies that at any given time low b-value regions will have more complete events than high b-value regions. Thus the minimum sample size should be determined with respect to the highest b-value (which is the slowest in accumulating complete events). Reciting the example from the previous section; two b-values of $b=1$ and $b=0.5$ will have completeness of %66 and %83 respectively. In another words equation (6) (with $k=3$) implies that by the time $b=1$ has accumulated 100 complete events $b=0.5$ will have 125 complete events. Thus $b=0.5$ will have more events and it should be considered with narrower confidence intervals. Accounting for this changing rate requires shifting the confidence plots of the lower b-values to the left with respect to the highest b-value. The normalization (shifting) coefficient ($C_i$) for a b-value $k'$ is basically the ratio of its completeness, given by equation (6), to the completeness of the highest b-value considered, $b_{max}$.

$$C_i = 1 - \frac{b_{max}}{k} \bigg/ 1 - \frac{k'}{k}$$
$$C_i = \frac{k - b_{max}}{k - k'}$$
(7)

Using the normalization coefficient we are able to calculate unbiased estimates of the minimum sample sizes required for different resolutions of $\Delta b$. Figure 3 shows the normalized confidence intervals and the obtained minimum sample sizes for $\Delta b$=0.16, 0.20, 0.24 and 0.28.

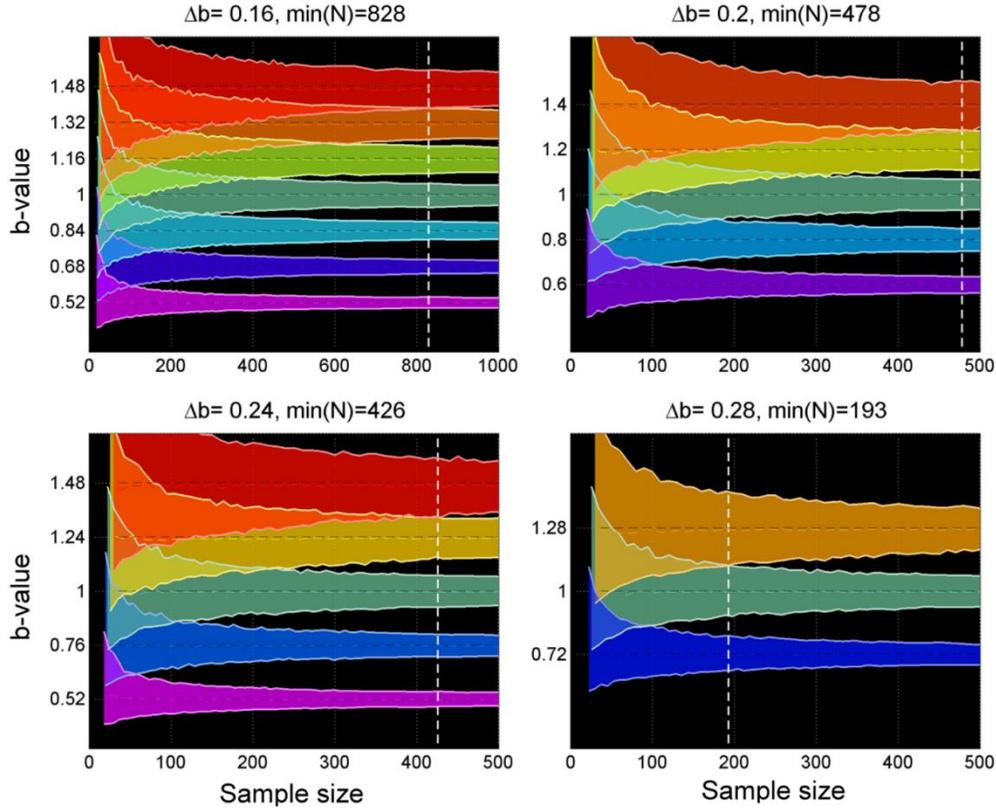

**Figure 3.** Normalized confidence intervals and corresponding minimum sample sizes for $\Delta b$=0.16, 0.20, 0.24 and 0.28 centered at $b$=1.

To convey the implications of our results more clearly we consider the lowest resolution which could be of scientific interest: the binary case of detecting low or high b-values. Assuming a regional b-value of $b$=1 and an interval of $0.5 \leq b \leq 1.5$ the lowest resolution to cover this interval would be $\Delta b$<0.5 since the coverage of the upper and lower bounds should also be considered. In Table 1 we tabulate the minimum complete sample size ($N$) and the detectable b-values at each resolution. The values in the middle rows represent the median, while the first and the third rows contain the 95% and 5% percentile. This implies that even a binary inference of high/low b-value requires a minimum of at least ~100 complete. On the other hand to be able to report a b-value variation with an acceptable standard deviation of at most 0.05 (~$\Delta b/2$) requires a sample size of at least ~500. Considering that many spatial and temporal b-value mapping studies have preferred to use sample sizes of 68 [*Tormann et al.*, 2013], 50 [*Öncel and Wyss*, 2000; *Wiemer and Wyss*, 2002; *Schorlemmer*, 2004; *Tormann et al.*, 2012], 40 [*Tassara et al.*, 2012] and even 25 [*Bachmann et al.*, 2012], the common skepticism regarding physical interpretations of b-value variation is not only justified but also necessary.

Table 1. Minimum number complete samples (N) required for resolutions of Δb. For each b-value column the 3 rows represent the 95% upper bound, median and %5 lower bound estimated over 1000 synthetic catalogs.

| N | b | | | | | | | | | Δb |
|---|---|---|---|---|---|---|---|---|---|---|
| | 0.6 | 0.7 | 0.8 | 0.9 | 1 | 1.1 | 1.2 | 1.3 | 1.4 | |
| 95 | 0.117 | | | | 0.200 | | | | 0.275 | |
| | 0.60 | | | | 1.00 | | | | 1.39 | 0.4 |
| | 0.095 | | | | 0.155 | | | | 0.232 | |
| 176 | | 0.097 | | | 0.132 | | | 0.175 | | |
| | | 0.70 | | | 1.00 | | | 1.29 | | 0.3 |
| | | 0.085 | | | 0.122 | | | 0.168 | | |
| 478 | 0.046 | | 0.061 | | 0.079 | | 0.095 | | 0.114 | |
| | 0.60 | | 0.80 | | 1.00 | | 1.19 | | 1.39 | 0.2 |
| | 0.042 | | 0.057 | | 0.071 | | 0.085 | | 0.104 | |
| 2345 | 0.020 | 0.024 | 0.027 | 0.029 | 0.034 | 0.038 | 0.040 | 0.043 | 0.049 | |
| | 0.60 | 0.70 | 0.80 | 0.90 | 1.00 | 1.09 | 1.19 | 1.29 | 1.39 | 0.1 |
| | 0.019 | 0.022 | 0.028 | 0.031 | 0.032 | 0.035 | 0.039 | 0.044 | 0.045 | |

The methodology described in this paper allows us to also estimate the minimum complete catalog required for a target resolution. Assume a target resolution $\Delta b$ with a minimum sample size of $N_{min}$. Using the normalization coefficient $C_i$ the sum of all complete events is given by equation (8) where $r$ is the number of all observable b-values.

$$N_c = \sum_i^r \frac{N_{min}}{C_i} \quad (8)$$

If we take for example the already tabulated $\Delta b=0.2$ ($N_{min}=478$ and $r=5$) $N_c$ is calculated as 2987. Thus by considering only the catalog size it is possible to rule out the presence of statistically significant b-value variation for a target resolution.

**Discussion and Conclusion**

In this study we tried to provide a generic answer to the question of the minimum sample size required for statistically significant detection of Gutenber-Richter's b-value. Using the angular FMD we conclude that the completeness ratio of the dataset changes with b-value. We have derived analytically a normalization coefficient to account for this effect in differentiating between varying b-values. Our results indicate that the minimum sample size used in many studies is almost an order of magnitude lower than the required value.

The presented results are fresh and rigorous as they account for both the incomplete and complete parts of the observed magnitude range and investigate a wide interval of b-value variation. However the problem of purporting power laws without statistical rigor is not new and is unfortunately widespread not only in statistical seismology. In their review article *Stumpf and Porter* [2012] showed that many power laws reported in various fields fail statistical testing. In the same paper they maintain that as a rule of thumb one needs a linear trend over at least two orders of magnitude on both *x* and *y* axes to claim existence of any power law, let alone distinguish between different exponents. It should be noted that this rule of thumb has been applied by some authors in their earlier studies [*Wiemer and Benoit*, 1996], however later on they have preferred to exclude it without any reminiscence.